\documentclass[%
 reprint,
superscriptaddress,
 amsmath,amssymb,
 aps,
]{revtex4-2}

\usepackage{graphicx}
\usepackage{dcolumn}
\usepackage{bm}
\usepackage{xcolor}
\usepackage{float}

\allowdisplaybreaks

\begin{document}

\preprint{APS/123-QED}
\title{Fast modeling of regenerative amplifier free-electron lasers}

\author{River~R.~Robles}
\email{riverr@stanford.edu}
\affiliation{Department of Applied Physics, Stanford University, Stanford, CA 94305.}
\affiliation{%
 SLAC National Accelerator Laboratory, Menlo Park, CA 94025.
}%
\author{Aliaksei~Halavanau}
\affiliation{%
 SLAC National Accelerator Laboratory, Menlo Park, CA 94025.
}%
\author{Gabriel~Marcus}\thanks{presently not at SLAC}
\affiliation{%
 SLAC National Accelerator Laboratory, Menlo Park, CA 94025.
}%
\author{Zhirong~Huang}%
\affiliation{Department of Applied Physics, Stanford University, Stanford, CA 94305.}
\affiliation{%
 SLAC National Accelerator Laboratory, Menlo Park, CA 94025.
}%

\date{\today}

\begin{abstract}

High-gain free-electron lasers (FELs) are becoming important light sources at short wavelengths such as the EUV and X-ray regimes. A particularly promising concept is the regenerative amplifier FEL (RAFEL), which can greatly increase the brightness and stability of a single pass device. One of the critical challenges of the x-ray RAFEL is maintaining electron-optical overlap over the relatively large (hundreds of meters) footprint of the system. Numerical modeling of x-ray RAFELs with angular and positional errors is critical for designing stable cavities, as well as to predict signatures of specific misalignment effects. Full-scale simulations of x-ray FELs are incredibly time-consuming, making large-scale parameter searches intractable on reasonable timescales. In this paper, we present a semi-analytical model that allows to investigate realistic scenarios - x-ray cavity without  gain (``cold cavity" or x-ray FEL oscillator) and x-ray RAFEL - in the presence of  angular/positional errors and electron trajectory oscillation. We especially focus on fast modeling of the FEL process and x-ray optics, while capturing effects pertaining to actual experimental setups at the Linac Coherent Light Source (LCLS) at SLAC. Such a method can be used to explore RAFEL at other wavelengths by suitable replacement of the optics modeling. 


\end{abstract}

\maketitle

\section{\label{sec:intro}Introduction}

Free-electron lasers (FELs) have revolutionized the scope of light-driven science. FELs are high-power, ultra-fast radiation sources which work by wiggling a relativistic electron beam inside of a magnetic undulator prompting the beam to emit synchrotron radiation. The central frequency of an FEL is tunable across the entire electromagnetic spectrum simply by changing the electron beam energy or the undulator field strength or period, allowing them to fill critical gaps left by conventional laser technology -- most notably in the THz, EUV, and x-ray spectral ranges. The earliest FELs operated at relatively long wavelengths, in the infrared and optical regimes \cite{hogan1998measurements,hogan1998measurements2,milton2000observation,milton2001exponential}. The conceptual designs of the earliest FELs looked very much like a traditional laser: a gain medium (the FEL) was enclosed within an optical cavity capable of storing the output radiation so that it could be amplified further and further with fresh electron beams \cite{madey1971stimulated,deacon1977first,nguyen1999first}. Similarly to conventional lasers, FELs of this form could be operated in a low gain-per-pass configuration (termed FEL osillator or FELO), or in a high gain-per-pass configuration (termed regenerative amplifier FEL or RAFEL). 

The true power of the FEL lies in its tunability, so the natural question is how to extend it to photon energies that are inaccessible by conventional laser technologies: x-rays and beyond \cite{murphy1985free,tatchyn1996research,pellegrini19932}. In this regime difficulties arose early on due to the lack of high reflectivity, large angle mirrors to form the optical cavity. As a result, the x-ray FEL (XFEL) has in every realization to date been a single-pass device, where the radiation is instantiated either by seeding with an external laser \cite{allaria2012highly} (only for very soft x-rays) or by amplifying the spontaneous radiation emitted by the electron beam due to its shot noise -- the latter is called self-amplified spontaneous emission (SASE) \cite{emma2010first,altarelli2011european,ishikawa2012compact,ko2017construction,milne2017swissfel}. SASE XFELs have been tremendously successful -- they are almost completely transversely coherent, but they suffer from a lack of temporal coherence  and shot-to-shot stability arising from the random nature of the startup \cite{huang2007review,pellegrini2016physics}. These issues limit the peak spectral brightness that can be obtained from SASE XFELs. Many concepts have been proposed to improve the temporal coherence, most notably self seeding \cite{amann2012demonstration,ratner2015experimental,emma2017experimental}, but for hard x-rays all are inherently limited by the fact that the start-up is always random. 

\begin{figure}[b]
    \centering
    \includegraphics{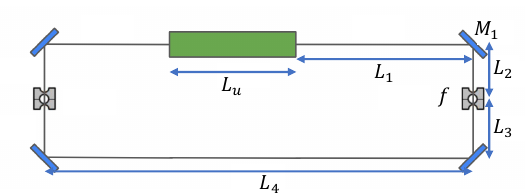}
    \caption{A possible example CBXFEL geometry. A rectangular cavity formed by four Bragg reflecting crystals wraps an undulator of length $L_u$. On either side of the cavity there are focusing lenses with focal length $f$. The mirror immediately following the undulator is labeled $M_1$.}
    \label{fig:cartoon}
\end{figure}

The discovery of very high ($\sim 99$\%) reflectivity Bragg reflections of hard x-rays from crystal structures \cite{liss2000storage,shvyd2003x,shvyd2011near} revitalized the discussions around cavity-based FELs in the x-ray regime \cite{huang2006fully,kim2008proposal}. An XFEL enclosed by a Bragg crystal cavity, which would provide both spectral and angular filtering to the radiation, promises to deliver fully coherent, high power, stable x-ray pulses. For later reference, Figure~\ref{fig:cartoon} shows the layout of a rectangular cavity-based FEL. As with the earliest FELs, cavity-based x-ray FELs (CBXFELs) are envisioned in two different regimes: the XFEL oscillator (XFELO) with low single-pass gain, and the x-ray regenerative amplifier FEL (XRAFEL) with high single-pass gain. There are several ongoing projects aimed at the first experimental demonstration of the CBXFEL concept at the LCLS (USA) \cite{marcus2019cavity}, EuXFEL (Germany) \cite{ PhysRevAccelBeams.26.020701} and SHINE (China) \cite{Huang2023} facilities. One of the key difficulties associated with the demonstration of CBXFELs is proper alignment of the optical cavity -- the XFEL gain medium has a 10s to 100 meter scale footprint while Bragg reflections have typical angular acceptances on the scale of 10 $\mu$rad. Thus these optical cavities must be aligned in principle with sub-$\mu$rad precision, and critically the effects of and tolerance to misalignment must be well understood.

In order to fully understand the tolerances and optimal design strategies for cavity-based FELs, accurate numerical modeling is essential. This is true at any wavelength, from the original IR cavity-based FELs, to extreme ultraviolet cavity-based FELs for lithography, to the x-ray regime. Cavity-based FELs pose a unique modeling challenge -- single-pass FEL simulations are notoriously time consuming and they must now be combined with Fourier optics propagation of the resulting fields through long cavities, and then all of this must be done over the tens to hundreds of passes required for the system to reach a steady state. In the low gain, FELO case, these challenges are alleviated by the fact that the impact of the FEL on the transverse dynamics of the photons is weak. One can thus largely get away with ignoring the impact of the FEL on the transverse dynamics, dealing only with the optical cavity \cite{PhysRevAccelBeams.25.090702,PhysRevAccelBeams.25.050701}. For RAFELs this is no longer the case, as high-gain FELs exhibit strong optical guiding effects -- both gain guiding and refractive guiding -- which play a dominant role in the dynamics of the optical pulse in the cavity \cite{prosnitz1981high,sprangle1987analysis,sprangle1987radiation,kim2017}. Thus, accurate modeling of RAFELs mandates accurate modeling of the FEL interaction, which makes simulations extremely time-consuming. This limits the extent to which the full RAFEL parameter space can be explored, holding back any individual project, and the field as a whole. Unfortunately, there is no real getting around how time consuming full fidelity RAFEL simulations are, however there is a space for faster models with more restricted accuracy which can be used to perform wide, coarse scans of the parameter space and study misalignment tolerances. The results found by such fast models could be then be used as starting points for finer scans with higher fidelity simulations.

In this paper we present a fast, semi-analytical model for RAFEL dynamics tailored precisely to that purpose. We base our technique on a time-independent, high-gain FEL gain guiding model combined with gaussian beam propagation. We start from the general 3D FEL equations which are applicable at any wavelength, but in our discussions we will focus on parameters relevant to hard XRAFEL systems. 

The paper is organized as follows. We first introduce gaussian beam parameterization and discuss the high-gain, time-independent FEL model as well as our approach to x-ray optics modeling. 
We then provide several concrete examples of where our technique can be applied, and use it for the case of an x-ray cavity with no gain (cold cavity), and a high-gain XFEL amplifier with and without e-beam trajectory oscillations. Finally, we utilize the unique capabilities of a fast modeling scheme to optimize the focusing system for an XRAFEL, and to understand the tolerance of a given XRAFEL design to increasingly large angular errors. 

\section{\label{sec:approach}Approach to fast cavity modeling}

Modeling cavity-based free-electron lasers requires dealing with two distinct phenomena: the first is the modeling of the FEL gain process in the undulator, and the second is propagation of the fields through the optical elements that make up the cavity. In this section we describe approximate methods for describing the propagation of a transversely gaussian beam through these two stages. In the FEL we restrict our attention to the high-gain regime (also called the exponential or linear amplification regime) before saturation, and for the cavity optics we restrict our attention to linear elements describable by ABCD matrices and introduce a simple model for Bragg reflections by crystals.

These approximations -- namely employing a gaussian ansatz and ignoring the saturation regime -- are motivated by the typical critical features of a RAFEL system and basic high-gain FEL physics. The high-gain FEL is characterized by strong gain and optical guiding effects which ensure a high degree of transverse coherence and a nearly gaussian transverse mode \cite{moore1985high,scharlemann1985optical,sprangle1987analysis,sprangle1987radiation,huang2007review,pellegrini2016physics,kim2017}. Even in the case of SASE, mode competition in the first few gain lengths gives rise, in the exponential gain regime, to a single gaussian mode that persists into early saturation. For an FEL amplifier initiated by a gaussian seed field -- which is an accurate description of all but the first pass in a RAFEL -- the field remains gaussian throughout the gain process. These facts justify the use of a gaussian ansatz for the sake of simplicity and speed. Ignoring the saturation regime is motivated more than anything by a lack of useful semi-analytical 3D models capable of capturing saturation effects. Thankfully, it is also in line with common RAFEL design methodologies which avoid allowing the FEL to go deep into saturation in a single pass. There are several reasons for that methodology. At saturation, gain and refractive guiding effects are reduced and the transverse mode quality suffers as a result \cite{huang2002transverse,saldin2003coherence,saldin2010statistical,geloni2010coherence}. Furthermore, the side-band instability causes the emission of unwanted frequency components \cite{kroll1981free,marcus2017x}. As we will find in our numerical examples later in the paper, for a well-designed RAFEL the linear theory is sufficient for finding good working points which can be further refined with traditional simulation tools. Conveniently, before saturation the FEL acts as a linear amplifier and does not couple different frequency components together. As a result, single frequency (or time-independent) modeling is sufficient to capture the essential physics. With all of these justifications in mind, we emphasize once more that the point of the presented model is to provide a fast tool for studying RAFEL systems so as to enable studies requiring a large number of simulations -- in particular, coarse parameter scans and tolerance studies. The ultimate verification of a given design should always be done with high fidelity numerical simulations.

We will employ the following monochromatic, gaussian ansatz for the field distribution
\begin{equation}\label{eqn:gaussianansatz}
    E(x,y,z) = f(z)e^{-\frac{i}{2}\left(Q_x(z)(x-x_0(z))^2+Q_y(z)(y-y_0(z))^2\right)}.
\end{equation}
This ansatz simplifies the problem both analytically and computationally by reducing the number of required parameters to describe the field to five. These parameters -- $f(z)$, $Q_x(z)$, $Q_y(z)$, $x_0(z)$, and $y_0(z)$ -- describe the field amplitude, mode size and divergence in $x$ and $y$, and transverse centroids in $x$ and $y$, respectively. In general they are all complex. We provide equations connecting these parameters to physically relevant ones in Appendix~\ref{app:physicalvalues}. In the rest of this section we will describe the evolution of these mode parameters through the optical elements relevant to XRAFEL operation. 

\subsection{\label{subsec:beamsinfibers}Gaussian beams in quadratic optical fibers}

We begin with a discussion of gaussian mode propagation through a quadratically gradient index optical fiber, which we will connect to the FEL propagation problem in the next subsection. The radiation profile in a material with some spatially varying refractive index $n(r,z)$ satisfies the following paraxial wave equation (see e.g. \cite{marcuse2013})
\begin{equation}\label{eqn:dielectricwaveguideeqn}
    \frac{\partial E}{\partial z} + \frac{1}{2ik_r}\nabla_\perp^2E = \frac{k_r}{2i}\left(1-n(r,z)^2\right)E,
\end{equation}
where $E$ is the slowly varying transverse field, $z$ is the propagation distance, and $k_r=2\pi/\lambda_r$ is the radiation wavenumber. For a quadratic gradient index fiber, we may write the refractive index (approximately or exactly, depending on the material), to second order in the transverse coordinates:
\begin{eqnarray}\label{eqn:secondorderindex}
    \nonumber n(x,y,z)^2 = &&n_0(z)^2+2n_{1x}(z)x+2n_{1y}(z)y\\
    &&-n_{2x}(z)x^2-n_{2y}(z)y^2.
\end{eqnarray}
The presence of the linear-order terms allows us to accurately take into account radiation profiles with non-zero transverse centroid. This form of the index couples naturally to our gaussian ansatz for the radiation field, as we can see from plugging the gaussian ansatz into equation~\eqref{eqn:dielectricwaveguideeqn} alongside equation~\eqref{eqn:secondorderindex}. The resulting equation is a second-order decoupled polynomial equation in $x$ and $y$: thus it has five terms that are proportional to unity, $x$, $y$, $x^2$, and $y^2$. We may take advantage of the orthogonality of the polynomial terms to separate the second-order polynomial equation in $x$ and $y$ into five distinct equations corresponding to each of the polynomial coefficients. The result is the following five first-order differential equations:
\begin{eqnarray}
    \label{eq:tracking}Q_x'(z) &=& \frac{k_r^2n_{2x}(z)+Q_x(z)^2}{k_r},\\
    Q_y'(z) &=& \frac{k_r^2n_{2y}(z)+Q_y(z)^2}{k_r},\\
    x_0'(z) &=& \frac{k_r(n_{1x}(z)-n_{2x}(z)x_0(z))}{Q_x(z)},\\
    y_0'(z) &=& \frac{k_r(n_{1y}(z)-n_{2y}(z)y_0(z))}{Q_y(z)},\\
    \nonumber f'(z) &=& \frac{f(z)}{2k_r}\left[Q_x(z)+Q_y(z)+ik_r^2(n_0(z)^2-1\right.\\
    \nonumber &&\left.+2n_{1x}(z)x_0(z)-n_{2x}(z)x_0(z)^2\right.\\ 
    \label{eq:tracking_end}&&\left.+2n_{1y}(z)y_0(z)-n_{2y}(z)y_0(z)^2)\right].
\end{eqnarray}
We note that this approach is analogous to that taken in \cite{kogelnik1965propagation} to study gaussian beam propagation through azimuthally symmetric gradient index fibers. The propagation of the gaussian mode can then be understood by integrating these equations through the system in question, given that we understand how to compute the refractive index components $n_0^2$, $n_{1x}$, $n_{1y}$, $n_{2x}$, and $n_{2y}$. Equations~\eqref{eq:tracking}-\eqref{eq:tracking_end} elucidate the primary reason for the fast nature of the method: we have reduced the problem of tracking the radiation profile down to tracking five complex numbers which obey simple first-order differential equations. 

\subsection{\label{subsec:felindex}The FEL as a quadratic optical fiber}

In order to make use of the formalism from the previous section, we must find a way to write the effective refractive index components of a high gain FEL. To describe the FEL interaction in the linear amplification regime with all relevant three-dimensional effects, we will borrow the formalism developed by Baxevanis \textit{et al} \cite{baxevanis2017}. As such, we consider here a planar undulator without any taper. Those authors showed that the FEL field development before saturation could be written concisely in the form of a single integro-differential equation
\begin{equation}
\label{eqn:panoseqn}
\begin{split}
    \frac{\partial E}{\partial z}&+\frac{1}{2ik_r}\nabla_\perp^2E = \\
    &\int dp_xdp_y\int_0^zd\zeta K_1(x,y,p_x,p_y,z,\zeta)E(x_+,y_+,\zeta),
\end{split}
\end{equation}
where $x$ and $y$ are the coordinates transverse to the undulator axis, and distance along the undulator is denoted by $z$. Furthermore, $p_x$ and $p_y$ describe the transverse propagation angles of a particle in $x$ and $y$. The variable $x_+=x\cos(k_\beta(\zeta-z))+(p_x/k_\beta)\sin(k_\beta(\zeta-z))$ with a similar expression in y. $k_\beta$ is the wavenumber associated with the beam betatron oscillations which we assume to be matched to an external smooth focusing lattice such that the beam divergence $\sigma_{x'}=k_\beta\sigma_x$, and the beam size and divergence are symmetric between the two planes and constant along the undulator. Additionally, the integral kernel $K_1$ has the form
\begin{widetext}
\begin{equation}
    K_1(x,y,p_x,p_y,z,\zeta)=K_{10}(\zeta-z)\exp\left[-\frac{(p_x^2+p_y^2+k_\beta^2x^2+k_\beta^2y^2)}{2k_\beta^2}\left(\frac{1}{\sigma_x^2}+ik_rk_\beta^2(\zeta-z)\right)\right].
\end{equation}
\end{widetext}
In this expression,
\begin{equation}
    K_{10}(\xi) = -\frac{8i\rho^3k_u^3}{2\pi k_\beta^2\sigma_x^2}\xi\exp[-i\Delta\nu k_u\xi-2\sigma_\delta^2k_u^2\xi^2],
\end{equation}
where $\rho$ is the Pierce parameter \cite{bonifacio1984collective}, $k_u=2\pi/\lambda_u$ is the undulator wavenumber, $\sigma_\delta$ is the relative energy spread of the electron beam, and $\Delta\nu$ is the FEL detuning parameter. Identifying the right-hand side of Equation~\eqref{eqn:panoseqn} with that of Equation~\eqref{eqn:dielectricwaveguideeqn} we may extract the effective local refractive index of the FEL 
\begin{equation}
\begin{split}
    n(x,y,z)^2 =& 1-\frac{2i}{k_r}\frac{1}{E(x,y,z)}\int dp_xdp_y\\
    &\times\int_0^zd\zeta K_1(x,y,p_x,p_y,z,\zeta)E(x_+,y_+,\zeta).
\end{split}
\end{equation}
In Appendix~\ref{app:index} we present an analytic form for this index for the gaussian field ansatz after the angular integrals have been taken, including the effects of a misaligned e-beam trajectory that we discuss further in the next section. The effective refractive index of a free-electron laser has been studied by multiple authors in the past \cite{prosnitz1981high,scharlemann1985optical,sprangle1987analysis,sprangle1987radiation} under different approximations, though never using a fully self-consistent formalism describing the coupled evolution of the electron beam and the radiation field as we are using. For the purposes of this work we would like to choose an appropriate approximation to put this index in the form of equation~\eqref{eqn:secondorderindex}. Although there are in principle many ways to do this, for the present case we will consider a simple Taylor expansion around the electron beam centroid. Physically, we expect this choice to capture the most relevant physics because the e-beam centroid is also the location of maximal gain, and the gain falls off exponentially with transverse offset. For radiation which is seeded far off-axis one expects that this will yield poor agreement in the early sections of the undulator, but, as long as there is sufficient total gain along the undulator length, we should expect good agreement by the undulator exit due to gain guiding. The accuracy of this relatively simple approach will be established quantitatively in the next section through numerical benchmarks. Thus we assign the approximate coefficients by 
\begin{eqnarray}
    n_0(z)^2 &=& n(x,y,z)^2\bigg\rvert_{x=y=0},\\
    n_{1x}(z) &=& \frac{1}{2}\frac{\partial n^2}{\partial x}\bigg\rvert_{x=y=0},\\
    n_{1y}(z) &=& \frac{1}{2}\frac{\partial n^2}{\partial y}\bigg\rvert_{x=y=0},\\
    n_{2x}(z) &=& -\frac{1}{2}\frac{\partial^2n^2}{\partial x^2}\bigg\rvert_{x=y=0},\\
    n_{2y}(z) &=& -\frac{1}{2}\frac{\partial^2n^2}{\partial y^2}\bigg\rvert_{x=y=0}.
\end{eqnarray}
These expressions can be evaluated analytically as a function of the local mode parameters such that the only integral left over is the integral in $\zeta$ which can be calculated numerically. These, coupled with the results of Subsection~\ref{subsec:beamsinfibers}, provide a self-consistent numerical scheme for propagating the seed radiation through the FEL. Extensive benchmarks of this approach are available in the earlier preprint \cite{robles2021self}.

Before moving on, we should note that within this exponential regime model, different frequency components propagate independently of one another. As a result, in principle the full bandwidth of the FEL can be simulated in parallel by tracking with different values for $\Delta\nu$. In this way, simple longitudinal effects such as bandwidth narrowing due to the finite mirror bandwidth can be taken into account. To account for more complex longitudinal effects, such as outcoupling schemes relying on electron beam shaping \cite{tang2022electron,tang2023active}, would require a revisiting of the underlying model and the basic single frequency approach may no longer apply. Including such effects is not critical to understanding basic transverse dynamics and stability of RAFELs as we explore in detail in later sections.

\subsection{\label{subsec:felwithorbits}The FEL optical fiber model with non-ideal beam trajectory}

Our model thus far assumes an idealized, on-axis e-beam transverse centroid trajectory. For real machine conditions, some level of shot-to-shot variation in the transverse centroid trajectory of the electron beam is unavoidable. This is important for XRAFEL operations, as for moderate levels of trajectory jitter the e-beam will continue to guide the radiation along its direction of motion leading to corresponding jitter in the pointing of the radiation at the exit of the undulator. For larger trajectories, the single-pass power gain can also be affected. Thus it is important to be able to take these effects into account in our model. For a smooth focusing lattice with wavenumber $k_\beta$ as we consider in our model, the trajectory given an initial transverse offset $x_{ce}(0)$ and angular offset $p_{x,ce}(0)$ is
\begin{eqnarray}
    x_{ce}(z) &=& x_{ce}(0)\cos(k_\beta z) + \frac{p_{x,ce}(0)}{k_\beta}\sin(k_\beta z),
\end{eqnarray}
with an equivalent expression in $y$, and $p_{x,ce}(z)=x_{ce}'(z)$.

The mechanics of allowing arbitrary trajectories in $x$ and $y$ are a straightforward extension of the previously presented formalism. We will list the few differences here, and in particular which expressions from before must be modified. To start, the original integral kernel is changed to
\begin{widetext}
\begin{eqnarray}\label{eqn:panoseqnarbtraj}
    \nonumber K_1(x,y,p_x,p_y,z,\zeta) &=& K_{10}(z,\zeta)\exp\left[-\frac{(p_x-p_{x,ce}(z))^2+(p_y-p_{y,ce}(z))^2}{2k_\beta^2\sigma_x^2}-\frac{(x-x_{ce}(z))^2+(y-y_{ce}(z))^2}{2\sigma_x^2}\right]\\
    & &\times\exp\left[-\frac{ik_r(\zeta-z)}{2}\left(p_x^2+p_y^2+k_\beta^2(x^2+y^2)\right)\right],
\end{eqnarray}
\end{widetext}
where we have introduced the predetermined electron beam centroid trajectories $x_{ce}(z)$ and $y_{ce}(z)$, in addition to the centroid momentum trajectories $p_{x,ce}(z)$ and $p_{y,ce}(z)$. In addition to this, we should now perform all Taylor expansions of the refractive index about the centroid of the electron trajectory, since that is now effectively the axis around which the most important physics occurs.

\begin{widetext}
\begin{equation}
    n(x,y,z)^2 = n_0^2(z)+2n_{1x}(z)(x-x_{ce}(z))+2n_{1y}(z)(y-y_{ce}(z))-n_{2x}(z)(x-x_{ce}(z))^2-n_{2y}(z)(y-y_{ce}(z))^2.
\end{equation}
\end{widetext}
Analogous to the case with no e-beam trajectory oscillations, the definitions of the various components of the refractive index are now modified to read, 
\begin{eqnarray}
    n_0(z)^2 &&= n^2\bigg\rvert_{x=x_{ce}(z),y=y_{ce}(z)},\\
    n_{1x}(z) &&= \frac{1}{2}\frac{\partial n^2}{\partial x}\bigg\rvert_{x=x_{ce}(z),y=y_{ce}(z)},\\
    n_{1y}(z) &&= \frac{1}{2}\frac{\partial n^2}{\partial y}\bigg\rvert_{x=x_{ce}(z),y=y_{ce}(z)},\\
    n_{2x}(z) &&= -\frac{1}{2}\frac{\partial^2 n^2}{\partial^2 x}\bigg\rvert_{x=x_{ce}(z),y=y_{ce}(z)},\\
    n_{2y}(z) &&= -\frac{1}{2}\frac{\partial^2 n^2}{\partial^2 y}\bigg\rvert_{x=x_{ce}(z),y=y_{ce}(z)}.
\end{eqnarray}
Finally, with this modified refractive index comes a requisite modified form for the evolution of the mode parameters:
\begin{eqnarray}
    Q_x'(z) &=& k_rn_{2x}(z)+\frac{Q_x(z)^2}{k_r},\\
    Q_y'(z) &=& k_rn_{2y}(z)+\frac{Q_y(z)^2}{k_r},\\
    x_0'(z) &=& \frac{k_r}{Q_x(z)}[n_{1x}(z)-n_{2x}(z)(x_0(z)-x_{ce}(z))],\\
    y_0'(z) &=& \frac{k_r}{Q_y(z)}[n_{1y}(z)-n_{2y}(z)(y_0(z)-y_{ce}(z))],
\end{eqnarray}
and the field amplitude now obeys, 
\begin{widetext}
\begin{eqnarray}
    \nonumber f'(z) &=& \frac{f(z)}{2k_r}\left[Q_x(z)+Q_y(z)+ik_r^2(-1+n_0^2+(x_0(z)-x_{ce}(z))(2n_{1x}(z)-n_{2x}(z)(x_0(z)-x_{ce}(z)))\right.\\
    &&\left.+(y_0(z)-y_{ce}(z))(2n_{1y}(z)-n_{2y}(z)(y_0(z)-y_{ce}(z))))\right].
\end{eqnarray}
\end{widetext}
Although the expressions have become more complicated, the basic numerical approach is unchanged. Furthermore, the results of Appendix~\ref{app:index} already include these extensions. 

\subsection{\label{subsec:cavityabcd}Optical cavity tracking}

The specific application of the XRAFEL demands that in addition to our approximate treatment of the FEL dynamics we also determine a way to track the radiation through the optical cavity. In conventional studies based on full FEL simulations, the field is stored on a cartesian grid, so the optical elements are likewise treated using a Fourier optics code. Here, in particular since all of the optical elements are simple drifts, lenses, and mirrors, we can simply use their well-known ABCD matrices. In the case of the drifts and lenses this is explicit, as it is known that when a gaussian beam propagates through an element described by an ABCD matrix it transforms according to the Huygens integral (see e.g. \cite{siegman1986lasers})
\begin{equation}
    E(x_f,y_f) = e^{-ik_rL}\int K_x(x_i,x_f)K_y(y_i,y_f)E(x_i,y_i)dx_idy_i,
\end{equation}
where $L$ is the length of the optical element along the optical axis and the integral kernels have the form 
\begin{equation}
    K_x(x_i,x_f)=\sqrt{\frac{i}{B\lambda_r}}\exp\left[-\frac{i\pi}{B\lambda_r}(Ax_i^2-2x_ix_f+Dx_f^2)\right],
\end{equation}
with a similar expression in $y$. The magnitude of the determinant of the ABCD matrix is one. For the gaussian beam these integrals result in another gaussian with mode parameters transformed according to  
\begin{eqnarray}
\label{eq:abcdQx0}
    Q_{x,y}&\rightarrow& k_r\frac{k_rC+DQ_{x,y}}{k_rA+BQ_{x,y}},\\
    x_0&\rightarrow&\frac{Q_xx_0}{k_rC+DQ_x},\\
    y_0&\rightarrow&\frac{Q_yy_0}{k_rC+DQ_y},\\
    \nonumber f&\rightarrow&\sqrt{\frac{k_r}{k_rA+BQ_x}}e^{-\frac{ik_rCQx_0^2}{2\left(k_rC+DQ_x\right)}}\\
    &&\times \sqrt{\frac{k_r}{k_rA+BQ_y}}e^{-\frac{ik_rCQ_yy_0^2}{2\left(k_rC+DQ_y\right)}}.
\end{eqnarray}
The approach described so far accounts for drifts and lenses which have 
\begin{equation}
    M_\mathrm{drift} = \begin{bmatrix} 1 & L \\ 0 & 1 \end{bmatrix}\hspace{1cm} M_\mathrm{lens} = \begin{bmatrix} 1 & 0 \\ -1/f & 1 \end{bmatrix}.
\end{equation}
The effects of loss in optical elements, for example air scattering in drifts and material losses in refractive lenses, can be accounted for by multiplying the amplitude by an appropriate loss coefficient. We note that for well-collimated beams and relatively weak focusing (the case pertaining to XRAFEL operation), the transverse beam size in the cavity remains small, well within the lens aperture. Therefore, the thin lens approach combined with a loss coefficient is an adequate model for intra-cavity beam focusing.

\subsection{Bragg diffraction of narrow divergence gaussian beams from perfect crystals}

\begin{figure}[h!]
    \centering
    \includegraphics{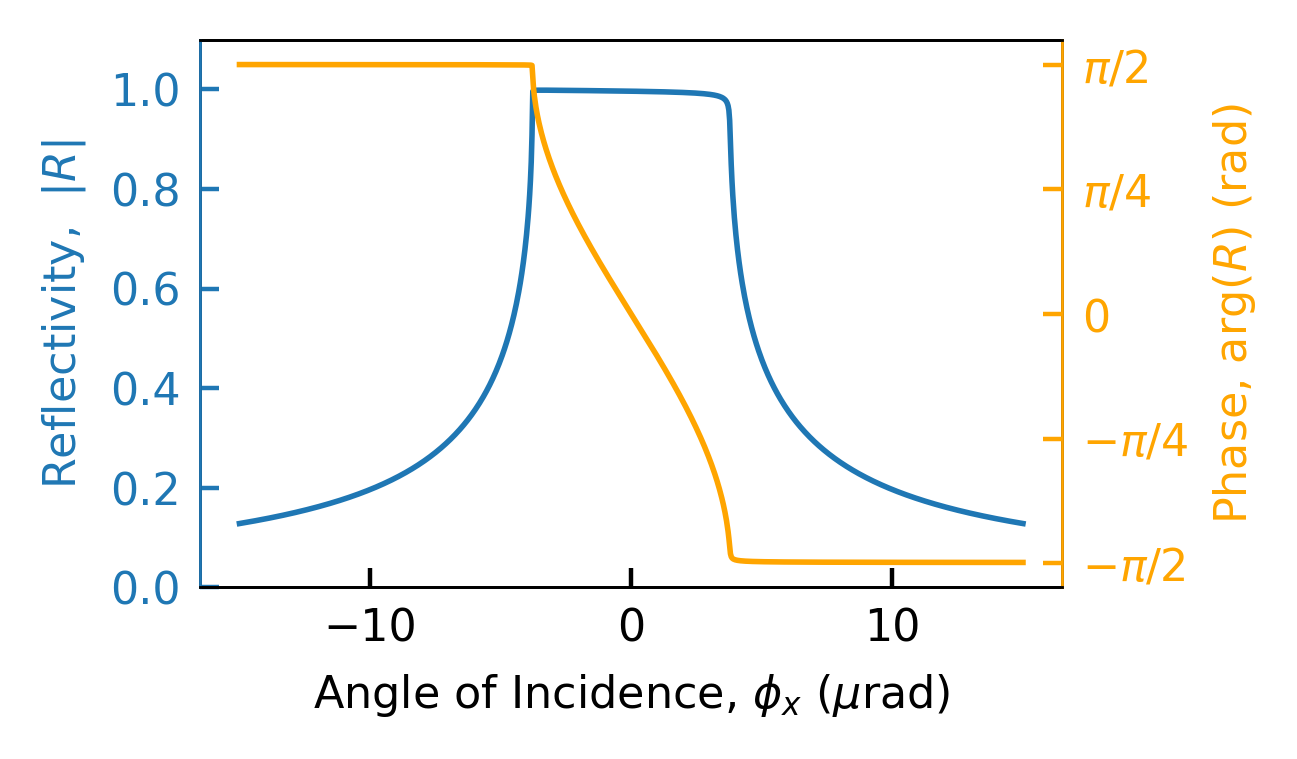}
    \caption{The magnitude and phase of the reflectivity curve for the $(4,0,0)$ reflection of 9.831 keV photons from diamond. The Darwin curve is re-centered by accounting for refraction with $\Delta \phi_r = -|\chi_0|/\sin{2 \phi_B}$. }
    \label{fig:reflectivity_C400}
\end{figure}

For x-ray RAFELs the mirror of choice is Bragg diffraction crystals. In general, the action of a Bragg diffraction crystal can be described by multiplying the spectral-angular representation of the field by a complex amplitude reflectivity function $R(\phi_x,\phi_y,\omega)$ which can be derived from dynamical diffraction theory \cite{Authier:399464} (in two-wave approximation). For a monochromatic field incident on a semi-infinite crystal, as we consider here, the amplitude reflectivity function (Darwin curve) can be written as:
\begin{equation}
    R(\phi_x,\phi_y) = \eta-\text{sign}[\Re[\eta]]\sqrt{\eta^2-1}\sqrt{\frac{|\chi_H|}{|\chi_{\bar{H}}|}},
\end{equation}
where $\eta$ is 
\begin{equation}
    \eta = \frac{-\phi_x\sin(2\phi_B)- \chi_0}{|P|\sqrt{\chi_H \chi_{\bar{H}}}},
\end{equation}
$\chi_0$ and $\chi_H$ here are the complex susceptibilities in the direction of incident and diffracted waves, $H$ is the reciprocal lattice vector, $\phi_B$ is the Bragg angle, and $P=1$ for $\sigma$ polarization and $\cos(2\phi_B)$ for $\pi$ polarization. Furthermore, $\phi_x$ is assumed to be measured relative to the Bragg angle. For reference we plot the magnitude and phase of the reflectivity curve for the $(4,0,0)$ reflection of diamond in Figure~\ref{fig:reflectivity_C400}. It has two primary features: a flat-topped region whose width is called the Darwin width, and slowly dropping Lorentzian tails. For an incident beam whose divergence is narrower than the Darwin width, the important parameters are the peak reflectivity and the slope of the linear phase shift across the flat-topped region. Note that the selection of angles by the Darwin curve is only present in the plane of diffraction (in our case $x$): the $y$ plane is unaffected by the reflection to first order. For this reason we delineate between the ``dispersive plane" and the ``non-dispersive plane". For the remainder of the paper we will use $x$ to refer to the dispersive plane.

In case of an incident gaussian beam, a Bragg diffracted beam is not strictly gaussian due to the nonlinear shape of the Darwin curve; see e.g. Refs. \cite{Authier:399464, PhysRevSTAB.15.050706,PhysRevSTAB.15.100702, Chu:76, McNeill:94}.
However, when the beam angular divergence is small, which is the case for XFEL radiation, the entire pulse stays within the Darwin width. For context, typical x-ray FEL divergences for hard x-rays are around or below 1 microradian \cite{huang2006fully,vartanyants2010coherence,parc2014radiation,turner:fel15-wep052}, whereas the typical angular acceptance of Bragg reflecting crystals at those same wavelengths are a factor of a few to an order of magnitude larger \cite{halavanau2023experimental} (e.g. see Fig.~\ref{fig:reflectivity_C400}). With this in mind we may say that for conditions relevant to XRAFEL operations - small FEL divergence, weak focusing in the large cavity, sub-$\mu$rad misalignments in the dispersive plane - the field is primarily acted upon by the flat portion of the Darwin curve and maintains its shape close to gaussian as a result. Similar assumptions have been previously made in the literature to study low gain cavity-based x-ray free-electron lasers; see, e.g. Refs. \cite{PhysRevAccelBeams.25.050701,PhysRevAccelBeams.25.090702}. Thus the key effects to take account of are the reduction of the field amplitude by the non-unity reflectivity coefficient and the linear phase shift across the Darwin width. We can capture these effects by defining an approximate reflectivity function as 
\begin{equation}
    R(\phi_x)\simeq R_0e^{ih\phi_x},
\end{equation}
where $R_0$ is a  reflectivity amplitude and $h$ is a phase coefficient that leads to the transverse shift. 

The final nuance to account for in the dispersive plane is that depending on the cavity geometry, the crystal acts either with $R(\phi_x)$ or with $R(-\phi_x)$. This ``dispersion sign" changes from one crystal to the next depending on the geometry. In a rectangular cavity, for example, the dispersion switches signs at each mirror, while in a bowtie cavity it only switches twice \cite{cotterill1968universal}. In the literature this sign changing is indicated with the notation $(+,+,+,+)$ for the rectangular cavity and $(+,-,+,-)$ for the bowtie: note that the sign in the notation is not the sign applied to the reflectivity function, it is just notation to indicate whether the sign locally flips or not. In the non-dispersive plane, the dynamics of a monochromatic wave are much simpler. The crystal acts like a perfect reflector in this plane, and so any angular misalignments of the crystal translate directly into angular misalignment of the reflected field. 
In this study we do not consider the finite thickness of the cavity mirrors for the sake of simplicity. We note that including a more rigorously derived formula of the crystal reflectivity (for a thin crystal) presented e.g. in Ref. \cite{PhysRevSTAB.15.100702}, will not change the results of our modelling.

We may represent the action of a misaligned crystal in the following way. First, the amplitude of the field after reflection with dispersion sign $\pm 1$ and angular offset in the dispersive plane $\phi_{m,x}$ becomes $\mathcal{E}_f(\phi_x,\phi_y,z)=R(\pm(\phi_x+\phi_{m,x}))\mathcal{E}_i(\phi_x,\phi_y,z)$. In this expression we use the definition of the angular field given in Appendix~\ref{app:physicalvalues}:
\begin{equation}
    \mathcal{E}(\phi_x,\phi_y,z) = \int E(x,y,z)e^{-ik_r(\phi_x x+\phi_yy)}dxdy.
\end{equation}
In addition to this complex amplitude change from the particular value of the crystal reflectivity, the x-ray pulse also experiences an angular pointing shift which is twice the mirror misalignment in the two transverse planes. Thus the field is subsequently shifted in $\phi_x$ by $2\phi_{m,x}$ and in $\phi_y$ by $2\phi_{m,y}$. In terms of our gaussian mode parametrization, the model reflectivity curve leads to the transformations
\begin{eqnarray}
    x_0&\rightarrow& x_0+\frac{2k_r\phi_{m,x}}{Q_x}\mp\frac{h}{k_r},\\
    y_0&\rightarrow& y_0+\frac{2k_r\phi_{m,y}}{Q_y},\\
    f&\rightarrow& fR_0e^{\pm i h\phi_{m,x}-2k_r\phi_{m,y}\Im[y_0]+\frac{2(k_r\phi_{m,y})^2\Im[Q_y]}{|Q_y|^2}}\\
    &&\times e^{-2k_r\phi_{m,x}\Im[x_0]+\frac{2(k_r\phi_{m,x})^2\Im[Q_x]}{|Q_x|^2}}.
\end{eqnarray}
In the tracking code used in the simulation studies below, we have implemented a flag to check whether the angular intensity of the field is close to the edge of the true Darwin curve. We do this by calculating the angular centroid of the radiation and adding to its absolute value 3 times the radiation divergence. If this value is greater than the half-width of the Darwin curve, a warning is shown and the simulation results are not trusted. In practice we will find that for well-designed XRAFEL systems this is not really a limitation.

\section{Cavity simulations with misalignments}

With our formalism now established, we will proceed to cavity simulations with comparisons between more conventional tracking methods and our approximate methods. We will focus on rectangular x-ray cavities (Bragg angle of 45$^\circ$) with the possibility to include focusing lenses. We focus on the photon energy 9.831 keV, which has a 45$^\circ$ Bragg reflection off of the $(4,0,0)$ plane of diamond that has gathered substantial interest from the CBXFEL community. Fitting to the flat portion of the Darwin curve (Figure~\ref{fig:reflectivity_C400}) gives the parameters $R_0\simeq 0.9965$ and $h=-0.3217$ rads/$\mu$rad for this reflection. Furthermore, the rectangular cavity exhibits a $(+,+,+,+)$ dispersion pattern (here $+$ means a change of dispersion direction). We note that it is important to properly account for the dispersion direction in the cavity, especially in the case when the radiation is focused before a mirror, and therefore ``inverted", with the lens. We further note that in this configuration, the volumetric shift of the transverse centroids by the crystals is perfectly cancelled by two adjacent crystals. The net effect of it is to increase the cavity length on the $\mu$m scale which can be trivially taken into account. It is important for establishing the correct timing overlap between the electron beam and the recirculating radiation pulse, but it has no impact on the transverse dynamics if lenses are shifted appropriately. Thus in the simulations below we artificially correct the shift after each mirror for both simulation methods.

To facilitate the benchmarking of our approach against more traditional methods, we have implemented an additional cavity tracking software which tracks the field on a square cartesian grid. The FEL interaction is handled using Genesis 1.3 Version 4 \cite{reiche1999genesis}. The Genesis simulations are performed in the time-independent, single frequency mode so as to enable direct comparison with our model. The optical cavity elements are tracked through using Fourier optics methods. The crystals are accounted for using the complete Darwin curve in the traditional tracking method.

\subsection{Cold cavity case}

Before dealing with the additional complexity of the FEL interaction, we begin by tracking a seed pulse through a ``cold" cavity: one without a gain mechanism. This allows us to observe the accuracy of our model for the optical elements in practice, most importantly the simplified crystal model. The parameters we choose are inspired by recent experimental work demonstrating the first low-loss ten-meter-scale x-ray cavity \cite{margraf2023low}. We consider a rectangular cavity with dimensions of 6 m by 1 m for a total round trip length of 14 m. In the middle of one of the 6 m sides we place an ideal thin lens with focal length 100 m. 

\begin{figure}[h!]
    \centering
    \includegraphics[width=0.9\columnwidth]{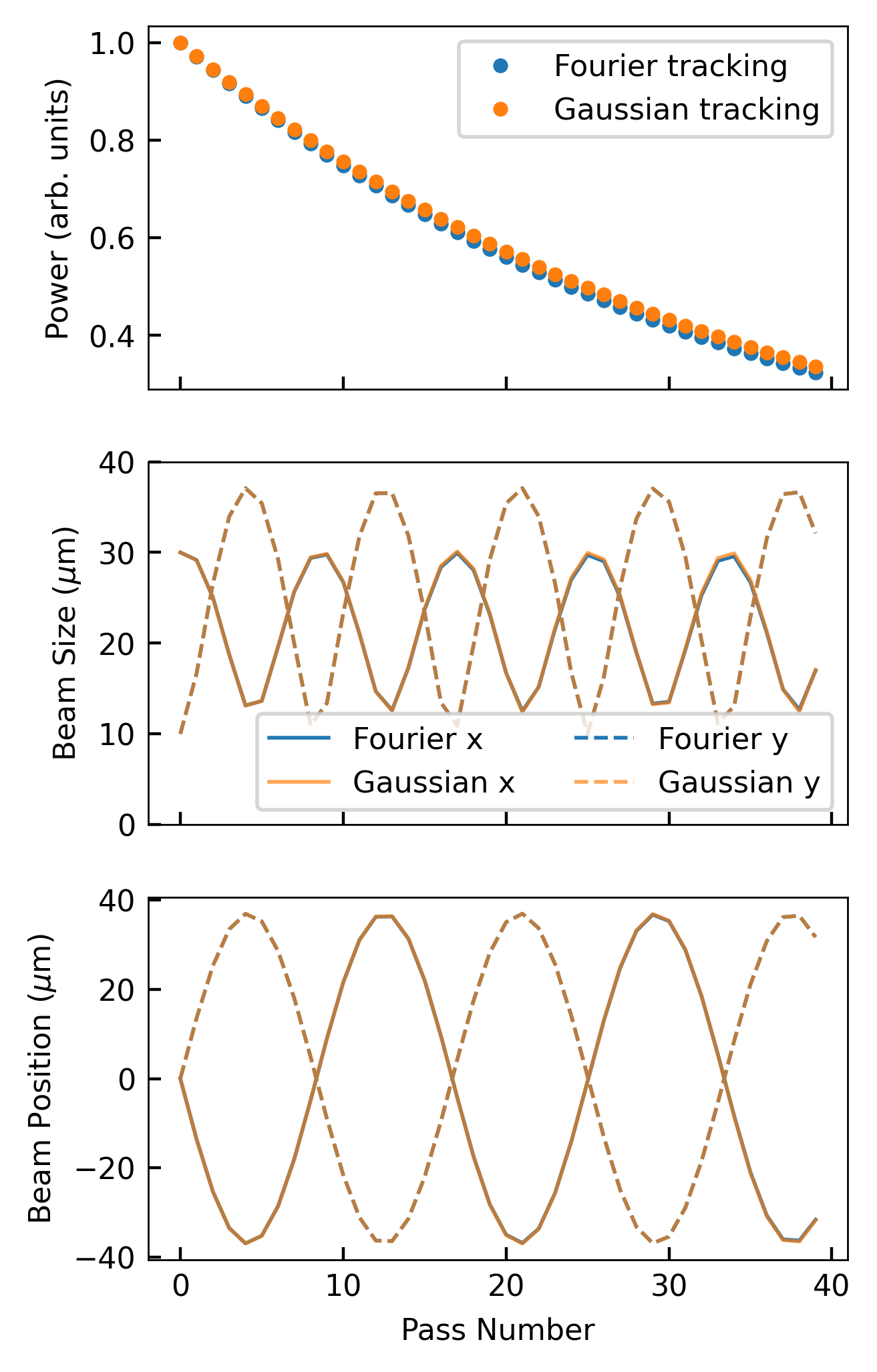}
    \caption{Tracking results for the cold cavity study. (a) The ringdown of the power stored in the cavity as a result of crystal losses. (b) The oscillations in the beam size in $x$ and $y$ due to combined effects of diffraction and focusing. (c) The oscillations in the beam centroids in $x$ and $y$ due to initial angular offset of the radiation.}
    \label{fig:coldcavity}
\end{figure}

We tracked a radiation pulse that is initially angularly displaced by 1 $\mu$radian in both planes, starting from a 30 $\mu$m waist in $x$ and a 10 $\mu$m waist in $y$. The traditional tracking method is done using a $\pm300$ $\mu$m grid spanned by 301 points. For each case we simulated 40 passes around the cavity: for the traditional method this took 11 seconds while for the Gaussian tracking method it took 25 milliseconds. We show the results of the simulations using both methods in Figure~\ref{fig:coldcavity}. Panel (a) shows the ringdown of intracavity power over the 40 simulated passes. On each pass there is a loss in power of roughly a factor $|R_0|^4$ due to the imperfect reflectivity amplitude of the Darwin curve. Slightly higher losses are observed in the Fourier tracking code because power leaks out of the tails of the radiation pulse that fall marginally outside of the Darwin width. Panels (b) and (c) present the transverse dynamics in the cavity, with the orange lines plotted with 70\% opacity to aid visibility. Due to the lens, the beam undergoes size oscillations in $x$ and $y$ with a frequency twice as fast as the oscillations in the beam centroids. The parameters we have chosen bring the beam size to a minimum value of 10 $\mu$m, which corresponds to a divergence of $816$ nrad. Furthermore, the maximum angular centroid offset in the dispersive plane is 1 $\mu$rad. This combination of angles means that the edges of the radiation field approach the edge of the Darwin curve, hence the very slight damping of the beam size oscillations in the dispersive plane which is not found in the Gaussian tracking code. This effect is quite small at this level because the angular centroid does not get too close to the Darwin width. This restriction is physically well-founded, for two reasons. First, when aligning an x-ray cavity there is a clear feedback on how well aligned the mirrors are in the dispersive plane since it directly talks to the storage efficiency of the cavity. For this reason, from here on we focus on mirror misalignments in the non-dispersive plane. Second, a well-designed XRAFEL system should not allow the beam divergence to become so large that it reaches the Darwin curve edges, with the exception of some outcoupling schemes \cite{PhysRevAccelBeams.26.020701, tang2023active}. Thus, we do not actually face much of a limitation from the flat crystal model, and our previously mentioned warning flag in the tracking code is sufficient to maintain high fidelity.

\subsection{High gain cavity case with static misalignments}

With the validity of our approach now established for basic optical elements, we move on to a rectangular cavity which wraps a high gain FEL. In addition to the losses due to the imperfect crystal reflectivity, we now also include losses due to absorption in the lens as well as potential outcoupling losses. We consider an FEL with parameters inspired by the planned LCLS-II-HE. The relevant FEL parameters are given in Table~\ref{tab:fel}. We now track the radiation field on a $\pm300$ $\mu$m grid spanned by 501 points.

\begin{table}[h!]
    \centering
    \begin{tabular}{c|c}
        \hline
        Beam parameter & Value  \\
        \hline
        Current & 1.5 kA\\
        Beta function & 20 m\\
        Norm. emittance & 0.4 $\mu$m\\
        Energy & 8 GeV \\
        Energy spread & 1.5 MeV\\
        \hline
        Undulator parameter & Value\\
        \hline
        Period & 2.6 cm \\
        Length ($L_u$) & 23.66 m\\
        Resonant photon energy & 9.831 keV\\
        \hline
        Cavity parameter & Value \\
        \hline 
        Short side length ($L_2/2=L_3/2$) & 1 m\\
        Long side length ($L_4=2L_1+L_u$) & 149 m
    \end{tabular}
    \caption{Parameters for XRAFEL simulations.}
    \label{tab:fel}
\end{table}

\begin{figure}[h!]
    \centering
    \includegraphics[width=0.9\columnwidth]{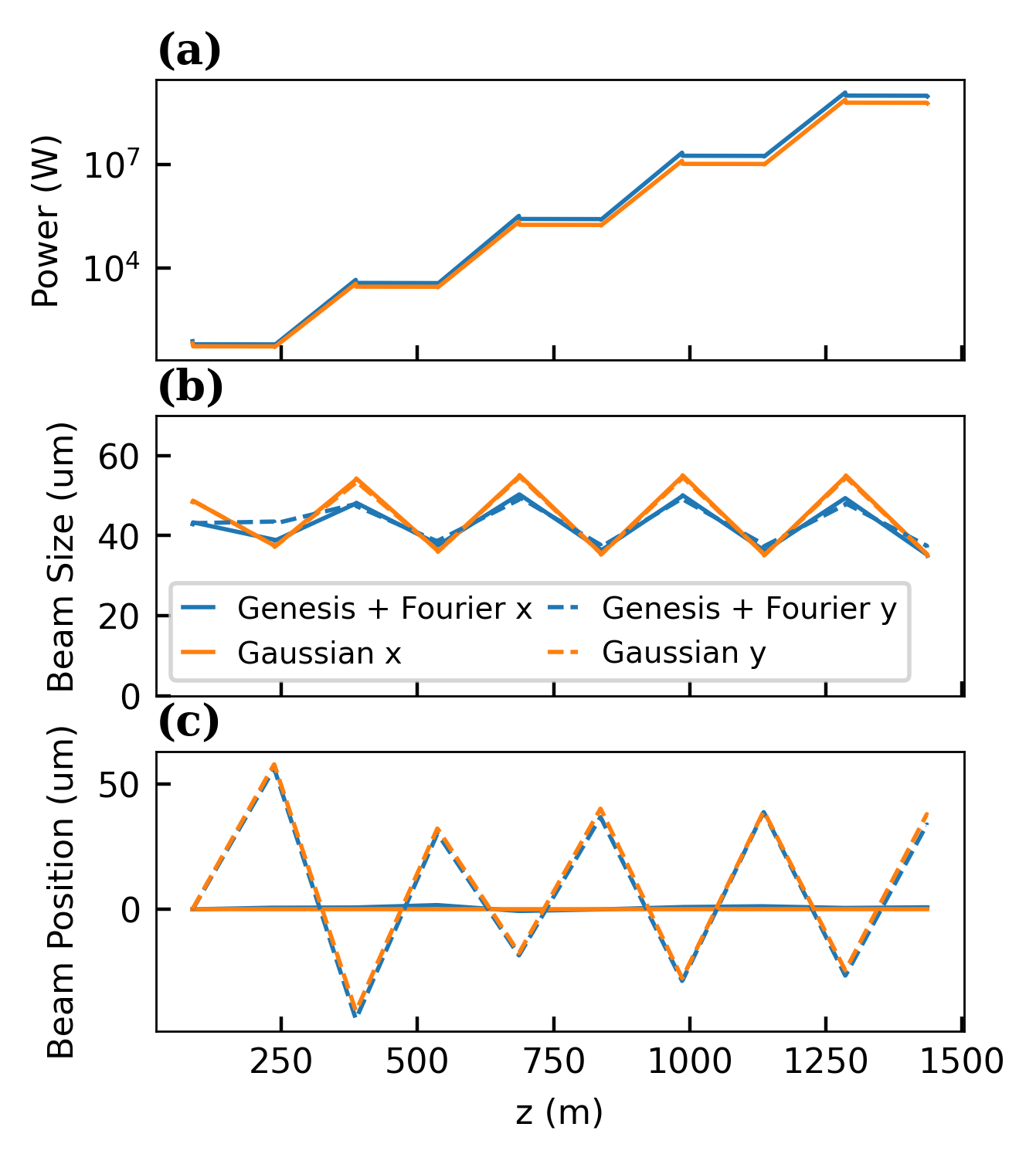}
    \caption{Tracking results for the cavity with gain study. (a) The power growth during propagation. (b) The oscillations in the beam size in $x$ and $y$ due to combined effects of diffraction and focusing from both the lens and the FEL. (c) The oscillations in the beam centroids in $x$ and $y$ due to initial angular offset of the radiation coupled with lens and FEL focusing.}
    \label{fig:cavity_with_gain}
\end{figure}

We now consider a rectangular cavity with 10\% energy outcoupling on each pass and two lenses of 50 meter focal length placed in the middle of the short drift sections in the cavity. The cavity geometry is described in Table~\ref{tab:fel}, where the included variable names correspond to the geometry shown in Figure~\ref{fig:cartoon}. We assume the lens to absorb 3\% of the radiation energy on each pass as well due to losses in the material. We take losses into account by simply multiplying the field by the appropriate factor. Additionally, as discussed previously it is in practice very difficult to exactly align the crystal angles in the non-dispersive plane. Thus, we pick at the start of the simulation four values from a normal distribution with 100 nrad rms spread and assign these values to the angular errors of the four crystals in the non-dispersive plane. We then simulate five cavity round trips with both simulation methods: the traditional method takes 1 minute while the gaussian method takes 2 seconds. Figure~\ref{fig:cavity_with_gain} shows the results of the simulations using the two methods, with the $x$ axis indicating physical distance along the optical path. Panel (a) shows the evolution of the radiation power, (b) shows the evolution of the rms beam size in $x$ and $y$, and (c) shows the beam positions in $x$ and $y$. Several interesting features are worth noting. In general the agreement between the two methods is not as exact as in the cold cavity case due to the complexity of the FEL interaction, but the results are still generally quite close. The beam size evolutions are slightly offset from each other by no more than 10 $\mu$m at any point as the grid-tracked pulse is not exactly gaussian. The centroid tracking does not suffer much from this, as the curves are almost perfectly overlapped in both planes. The centroids exhibit oscillations in $y$ due to the coupling of angular misalignments to both the focusing from the lens as well as the optical guiding in the FEL. 

\subsection{High gain cavity case with e-beam trajectory oscillation}

\begin{figure}[h!]
    \centering
    \includegraphics[width=0.9\columnwidth]{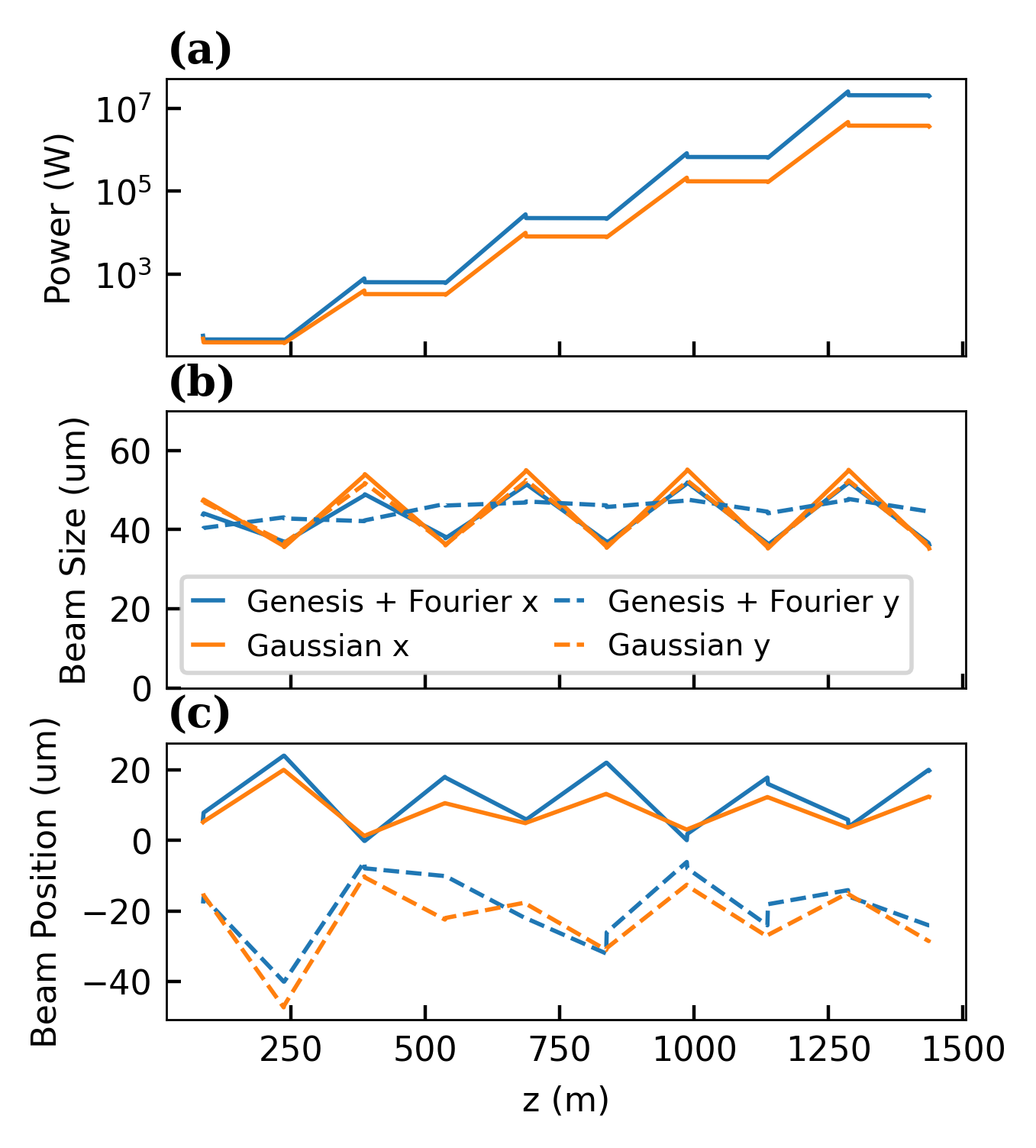}
    \caption{Tracking results for the cavity with gain and e-beam trajectory study. (a) The power growth during propagation. (b) The oscillations in the beam size in $x$ and $y$ due to combined effects of diffraction and focusing from both the lens and the FEL. (c) The oscillations in the beam centroids in $x$ and $y$ due to initial angular offset of the radiation coupled with lens and FEL focusing.}
    \label{fig:cavity_with_gain_and_orbit}
\end{figure}

Electron beam trajectory oscillations are unavoidable features of XFEL linacs. Although they can be minimized, they can never be completely eliminated. A misaligned e-beam in the FEL undulator leads to deleterious guiding of the seed radiation away from the nominal optical axis for small trajectories, and for large ones it can even lead to degradation of gain. To ensure that our fast tracking model can account for these complicated but important effects, we performed cavity simulations as in the last section, but this time we initiated the beam with -0.5 $\mu$rad and 1.5 $\mu$rad initial angles in the $x$ and $y$ directions, respectively. Similarly, as before the mirrors were given initial random misalignments in the non-dispersive plane. Figure~\ref{fig:cavity_with_gain_and_orbit} shows the results of that study in the same style as Figure~\ref{fig:cavity_with_gain}. Although the dramatic entrance angles lead to an underestimation of the instantaneous power, the transverse dynamics are still very well represented by the fast cavity model. We also note that for this particular configuration, the errors conspire to shift the effective optical axis away from the nominal rectangular one. This can be seen in panel (c), in which the centroid seems to oscillate not around the axis, but rather some perturbed value. This concept has been discussed largely for low gain XFEL oscillators, in which case the distortion of the optical axis is due to optical element misalignments \cite{KJKtechnote}. This case shows that errors in the FEL can similarly distort the optical axis in a high gain configuration. A recent study found similar, though smaller amplitude, results for the low gain case \cite{PhysRevAccelBeams.25.090702}.

\section{Numerical studies using the fast code}

We now proceed to demonstrate the real advantage of having a sufficiently accurate, fast tracking code for the cavity: fast scanning of cavity parameters. Performing cavity parameter scans on a large scale is extremely time-consuming using traditional methods due to the long single simulation times. This is especially problematic for statistically varying quantities like mirror misalignment angles and electron beam trajectories, for which one must run many simulations with randomly sampled errors. Here we will use the fast cavity model to study two extremely important topics in RAFEL design: the proper choice of intra-cavity focusing strengths, and the impact of mirror misalignments. 

\subsection{Intra-cavity focusing optimization}

The placement and strength of the intracavity lenses must both be highly optimized in order to ensure long-term cavity stability and maximal cavity output power. Traditionally, the focal length for a given lens configuration has been picked to set up a radiation waist in the middle of the undulator in the equivalent cold cavity scenario. This approach makes sense for low gain systems, but is not necessarily ideal for high gain RAFEL cavities in which the FEL itself is a non-trivial optical element that contributes to the cavity transverse dynamics. Thus, as a first application of the method we will utilize the fast modeling scheme to scan lens focal lengths to determine an optimal cavity performance working point taking into account FEL guiding. For both studies shown below we utilize the same basic cavity geometry as in the benchmarks with two lenses of the same focal length on the two short sides of the cavity.

\begin{figure}[h!]
    \centering
    \includegraphics[width=\columnwidth]{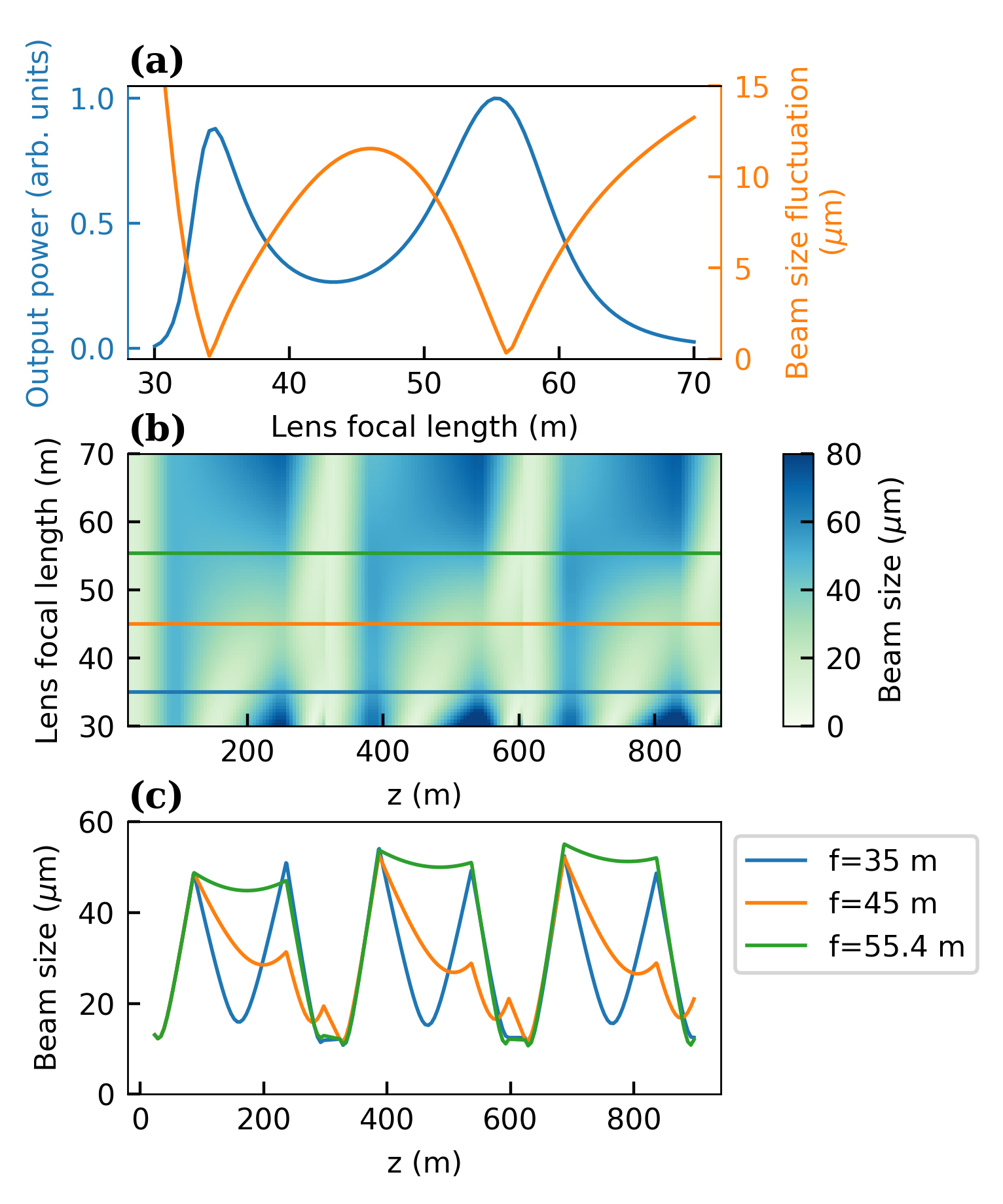}
    \caption{(a) The output power (normalized to the maximum value) and the root-mean-square fluctuation in beam size over the last two passes are plotted as the focal length of the lenses is scanned. (b) A sweep through focal length values as a function of intracavity beam size (in color). (c) The beam size evolution over three cavity passes is shown for three characteristic focal lengths.}
    \label{fig:lens_scan}
\end{figure}

Figure~\ref{fig:lens_scan} shows the results of a scan of the lens focal lengths from 30 m to 70 m. Panel (a) shows the output power after 5 passes normalized to the maximum in the scan range, alongside the rms fluctuation of the beam size on the four crystals in the last two passes. There are peaks in the output power for $f=35$ m and $f=55.4$ m coinciding with the minimization of the beam spot size fluctuation on the crystals. We note that the flag on angular width we mentioned previously begins to give warnings for focal lengths smaller than $32$ m. Panel (b) shows the oscillation of the beam size in the cavity over 3 round trips as the lens focal length is scanned. The two optimal points coincide with symmetric, periodic solutions and deviation on either side removes the symmetry within a single pass. Panel (c) shows the oscillation of the radiation beam size over 3 passes through the cavity for three characteristic focal lengths: the two at which the output power peaks $f=35$ m and $f=55.4$ m, and the local power minimum between them $f=45$ m. In Panels (b-c), $z=0$ corresponds to the exit of the FEL on the first pass and the final point near $z=900$ m corresponds to the entrance of the FEL starting the fourth pass. From these oscillations we see that the optimal focusing configurations correspond to establishing symmetric oscillations of the beam size through the cavity, with both a small and a large radiation mode in the middle of the long side of the cavity being sufficient as long as the beam is focused near the entrance of the undulator. For the non-optimal focal length of $f=45$ m, the beam is larger entering the undulators which reduces the gain of the radiation field. 

\begin{figure}[h!]
    \centering
    \includegraphics[width=1\columnwidth]{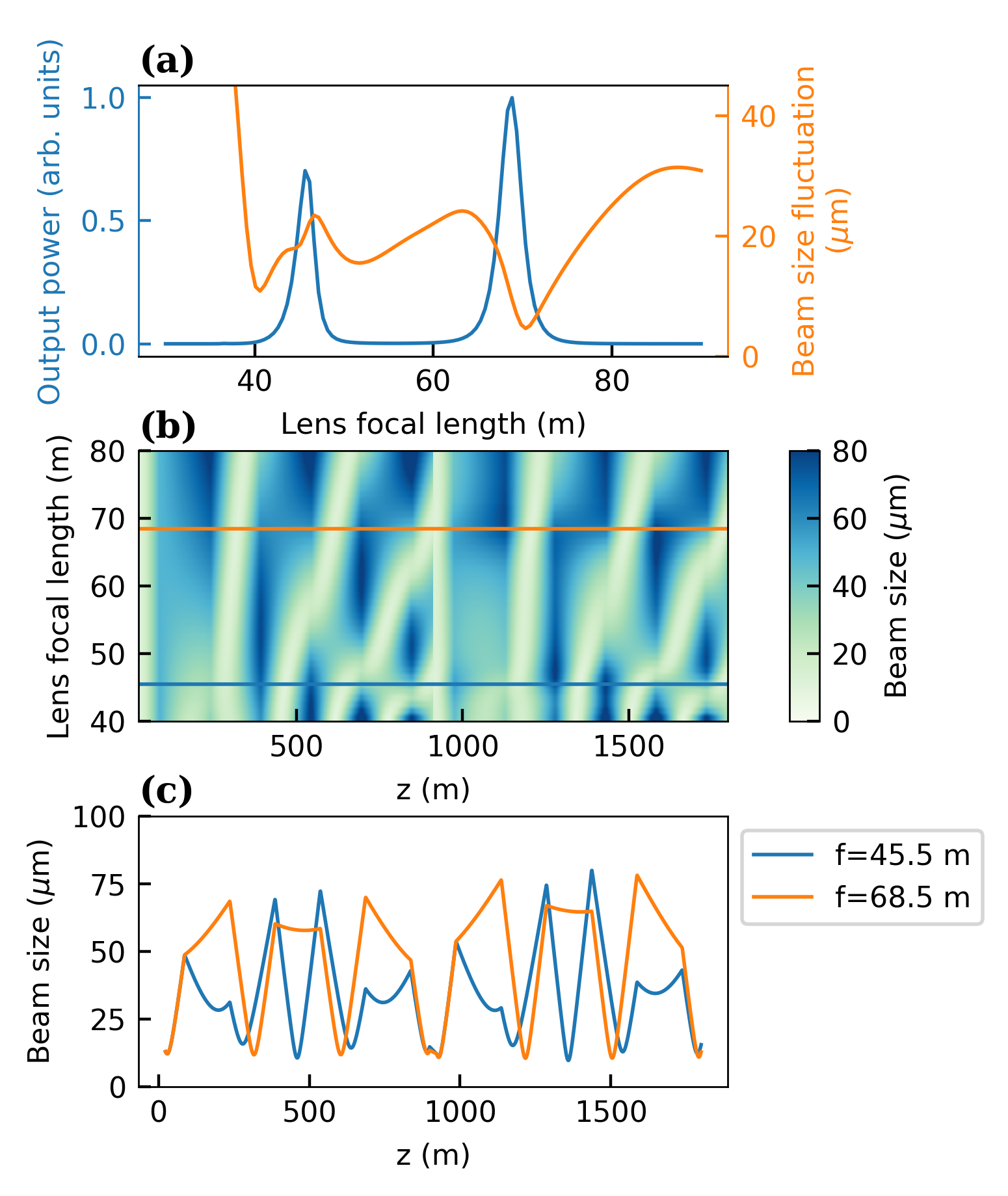}
    \caption{(a) The output power (normalized to the maximum value) and the root-mean-square fluctuation in beam size over the last two passes are plotted as the focal length of the lenses is scanned. In this case the field only interact with an electron beam on every third pass through the cavity. (b) A sweep through focal length values as a function of intracavity beam size (in color). (c) The beam size evolution over three cavity passes is shown for two characteristic focal lengths.}
    \label{fig:lens_scan_threepass}
\end{figure}

So far we have assumed that the theoretical linac repetition rate perfectly matches the cavity round trip time such that we have amplification in the FEL on every round trip. This does not generally need to be the case, and indeed in some designs it is necessary to let the radiation propagate through the cavity over multiple passes without interacting with an electron beam \cite{tang2023active}. Thus for another study we consider a linac with a 333 kHz repetition rate such that the radiation interacts with an electron beam on every third round trip. We expect since the interaction with the e-beam is an essential component of the cavity transverse dynamics that the optimal focusing location will change. Indeed, Figure~\ref{fig:lens_scan_threepass} shows the equivalent to Figure~\ref{fig:lens_scan} for this new repetition rate configuration. This time, the warning flag is shown for focal lengths smaller than $40$ m. Panel (a) again shows two peaks in the output power, this time for $f=45.5$ m and $f=68.5$ m. The shift of the two optimal focal lengths relative to the prior study is testament to the strong impact of the FEL on the transverse dynamics in the cavity, and the necessity of accounting for FEL optical guiding when designing XRAFEL systems. We note that a similar conclusion has been reported in Ref. \cite{10.1063/1.5037180}. Additionally, the tolerance on the lens focal length is tighter for this configuration: whereas the full-width at half-maximum for the output power in the previous study was nearly 10 m for the $f=55.4$ m working point, here it is much narrower. Panel (b) shows the beam size oscillations for focal lengths between 40 m and 80 m, while Panel (c) shows the particular oscillations for the two optimal focal lengths over six round trips. The last point again coincides with the entrance of the undulator. The optimal focal lengths bring the field to a focus well-matched to the natural FEL mode side just before the entrance of the undulator around $900$ m and $1800$ m.

\begin{figure*}[htb!]
    \centering
    \includegraphics{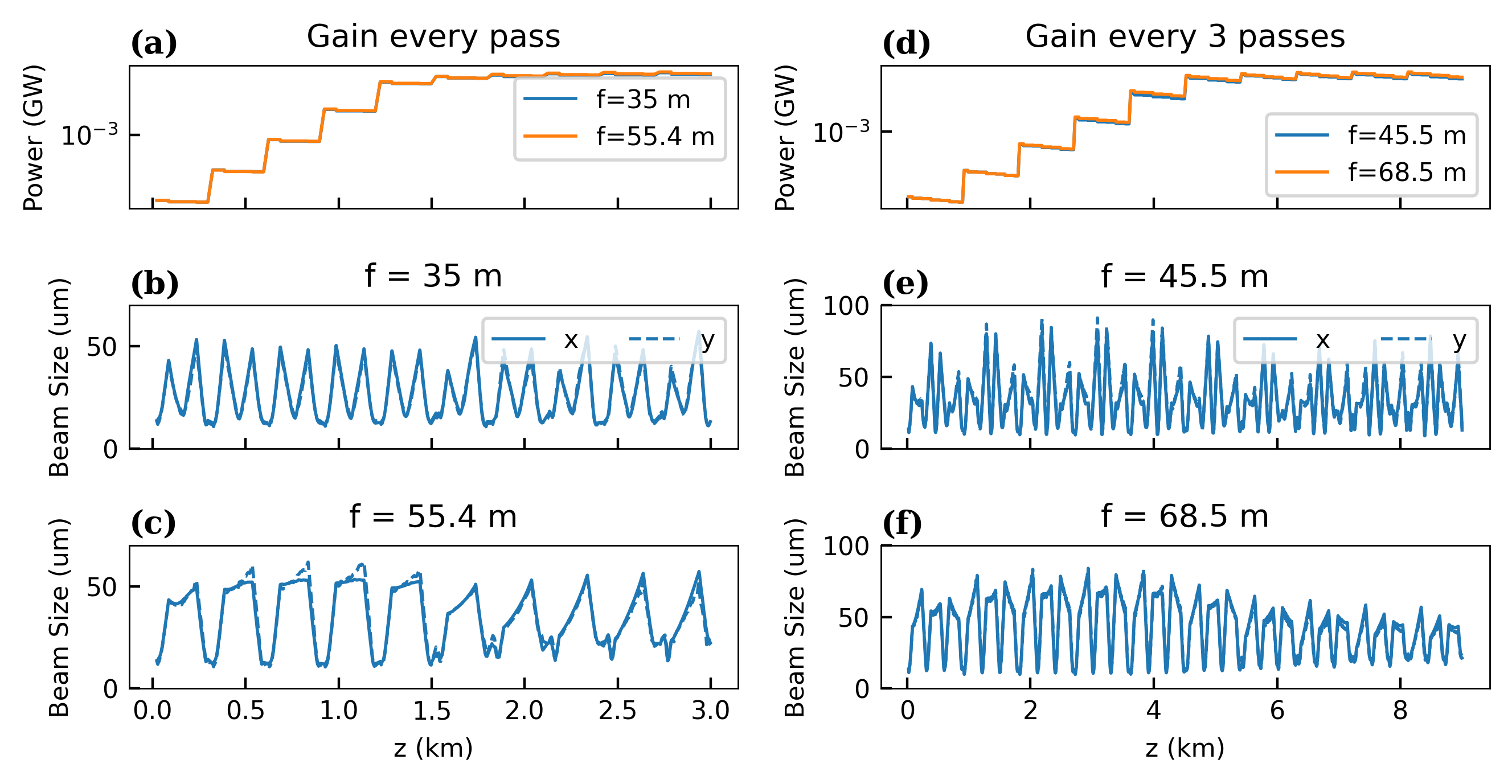}
    \caption{XRAFEL simulations using traditional methods of a rectangular cavity with optimal focusing as predicted by the fast code. The top row shows power evolution through the cavity and the bottom two rows show oscillations in the rms beam size in both planes. (a-c) A cavity with FEL gain on every pass with either a 35 m (b) or 55.4 m (c) focal length. (d-f) A cavity with FEL gain every three passes with either a 45.5 m (e) or 68.5 m (f) focal length.} 
    \label{fig:genesis_comparisons}
\end{figure*}

The working points discussed above can act as starting points for finer, more accurate optimization using traditional methods. It is important to note that our method does not account for the saturation regime of the FEL, during which optical guiding effects are reduced. To further validate our claim that these configurations found by the fast code are still feasible when the XRAFEL is driven to its saturated steady state, we have simulated the above XRAFEL systems using traditional methods for the optimal focal lengths. Figure~\ref{fig:genesis_comparisons} shows the results of these simulations. Panels (a-c) deal with a cavity with FEL gain on every pass as we studied in Figure~\ref{fig:lens_scan}. The top row shows the power as the field propagates through 10 passes of the cavity. The bottom two rows (panels b and c) show the beam size oscillations in both $x$ and $y$ for the two lens focal lengths. The FEL reaches the saturation regime around 1.3 km of propagation in both simulations. We note that the beam size oscillations before saturation match very well with the corresponding lines in Figure~\ref{fig:lens_scan}, panel (c) as expected. We observe very little qualitative change in the beam size oscillations with 35 meter focal length (panel b) as the FEL transitions from the linear to the saturated regime. On the other hand, the 55.4 meter focal length scenario changes somewhat significantly as we pass into the saturation regime. This can be understood by looking at the differences between the two oscillations in the linear amplification regime. The 35 meter cavity exhibits oscillations that almost look like oscillations in a cold cavity: the beam size is roughly symmetric between the two long sides of the cavity despite the presence of the FEL on one of the sides. Thus, the reduction of the FEL guiding effects during saturation has a relatively small impact on the oscillations. The 55.4 meter focal length case exhibits asymmetric oscillations between the two long cavity sides that are only maintainable due to the FEL guiding. A similar observation comes from looking at panels (d-f), where we study an XRAFEL that has gain only every three round trips as in Figure~\ref{fig:lens_scan_threepass}. Now the saturation point is pushed back to roughly 5 km of propagation due to the less frequent FEL gain. As with the previous study we see that the shorter focal length working point maintains qualitatively similar characteristics into saturation compared to the longer focal length working point. It is also worth keeping in mind that the larger focal length operating points keep the the radiation divergence smaller on the crystals, which may be desirable for stability when mirror jitter is taken into account. In all of these cases, the goal of the fast model -- finding a reasonably accurate starting point for fine tuning -- has clearly been met, and additionally it has unveiled interesting physical effects in the cavity dynamics.

\subsection{XRAFEL tolerance to static mirror misalignments}

\begin{figure}[h!]
    \centering
    \includegraphics[width=0.9\columnwidth]{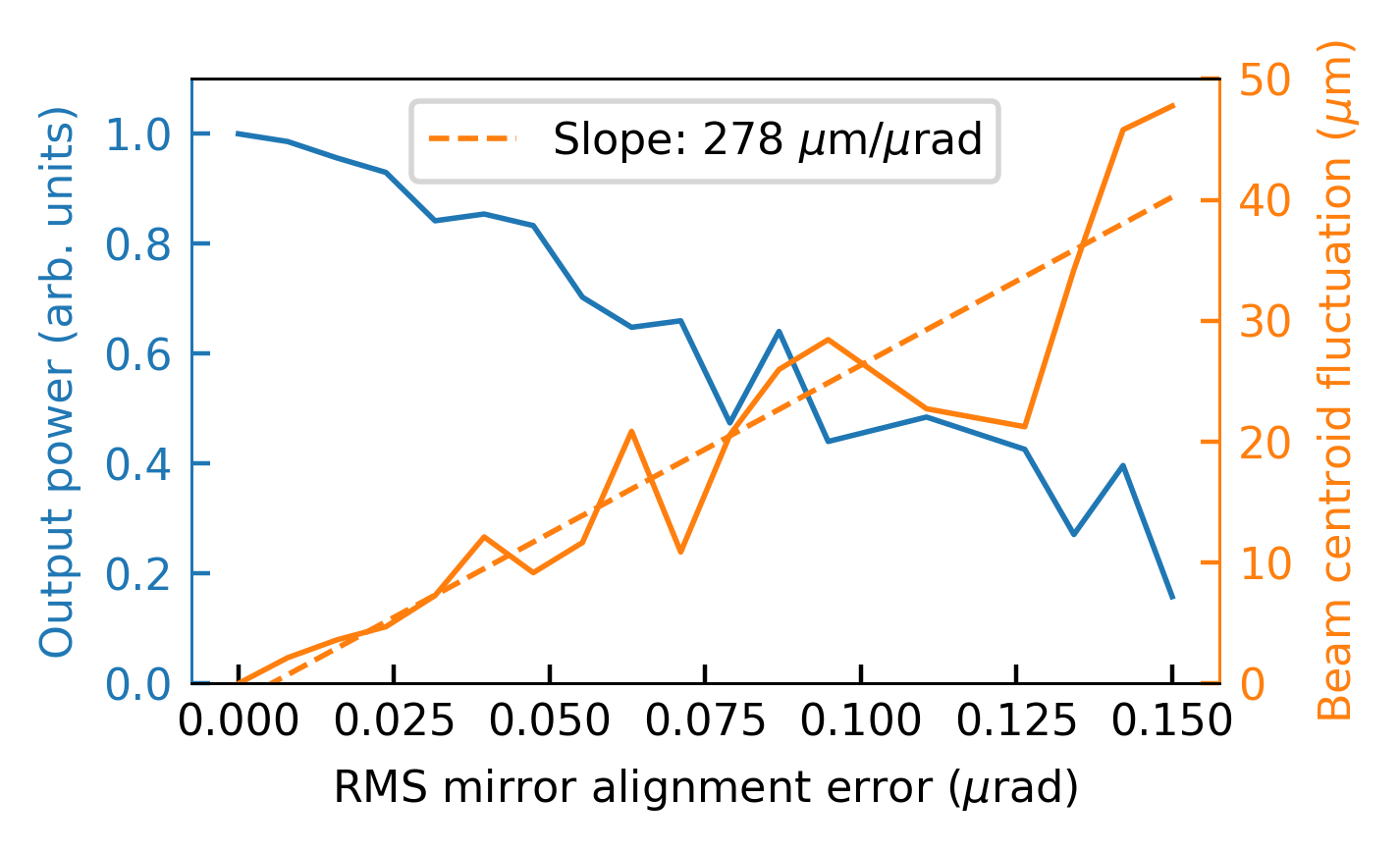}
    \caption{The output power (normalized to the maximum value) and the root-mean-square fluctuation in beam centroid on the last mirror are plotted as the rms mirror misalignment angles are increased. Each datapoint is obtained by averaging over 10 statistically independent simulations. }
    \label{fig:jitter_scan}
\end{figure}

In addition to determining the optimal cavity design for otherwise perfectly aligned elements, it is critical to understand the behavior of a particular cavity design in the presence of errors. Here we focus on static mirror misalignments in the non-dispersive plane. We consider that the cavity alignment is not perfect, in particular because the non-dispersive plane is much more difficult to align to high precision than the dispersive one. As a result, the mirrors may be left with some fixed, non-zero misalignment angles. We sample the misalignment of the four mirrors from a normal distribution with a variable standard deviation, and repeat this 10 times for each standard deviation in order to gather stable mean values for the behavior of the output power and radiation pointing jitter. Figure~\ref{fig:jitter_scan} shows the output power and beam centroid fluctuation on the crystal immediately following the undulator ($M_1$ in Figure~\ref{fig:cartoon}) as a function of this rms mirror alignment error. We see that with 150 nrad mirror misalignment the output power of the cavity has dropped to 20\% of its perfectly aligned value. Furthermore, the centroid of the radiation on $M_1$ jitters with a standard deviation that scales roughly linearly with the rms mirror misalignment. The slope of this linear dependence was found to be 278 $\mu$m/$\mu$rad, meaning that 1 $\mu$rad rms misalignment error would lead to a 278 $\mu$m rms jitter of the radiation centroid on $M_1$. This study informs the required alignment precision in the non-dispersive plane for such a cavity geometry.

\section{\label{sec:conclusions}Conclusions}

We have presented a fast modeling methodology for regenerative amplifier FELs which leverages the nearly gaussian shape of the field propagated through highly optimized RAFEL systems. Our scheme enables rapid testing of different RAFEL configurations and predicts nearly the same results as traditional tracking codes as long as the beam divergence and pointing do not approach the Darwin width. This assumption is reasonable, as a well-designed XRAFEL system should not approach the Darwin width anyway. We have shown how our fast tracking code can be used to optimize parameter values in XRAFEL designs with the specific example of the lens focal length. This includes designs that are robust up to the saturation regime of the XRAFEL, so that our preliminary workings points act as good starting points for finer tuning with higher accuracy simulations. Furthermore, the fast tracking method can be used to understand the behavior of a given XRAFEL design in the presence of misalignments and electron beam trajectory oscillations, which is critical for ensuring a stable, reliable working point. 

\section*{Acknowledgements}

This work was supported by DOE Contract DE-AC02-76SF00515. R.R.R. acknowledges support from the William R. Hewlett graduate fellowship through the Stanford Graduate Fellowship (SGF) program as well as the Robert H. Siemann Fellowship. We are grateful to Diling Zhu (SLAC) for many insightful comments on the subject.  

\appendix

\section{\label{app:physicalvalues}Conversion to physical values}

To facilitate comparison of our results with simulations we must convert the complex radiation mode parameters we are tracking into measurable quantities. The parameters of the most interest are the radiation rms or fwhm size, physical centroid, angular centroid, and power. We will consider the transverse sizes and centroids of the transverse intensity profile $I(x,y,z)=\frac{\epsilon_0c}{2}|E(x,y,z)|^2$. The rms size is 
\begin{equation}
    \sigma_{x,y} = \sqrt{-\frac{1}{2\Im[Q_{x,y}]}}.
\end{equation}
Since the beam is gaussian the full-width at half-maximum size is related to this by fwhm $=2\sqrt{2\log(2)}\sigma$. Furthermore the physical centroid, which we identify as the location of the peak of the intensity profile, is 
\begin{equation}
    x_{cen} = \Re[x_0]+\Im[x_0]\frac{\Re[Q_x]}{\Im[Q_x]}.
\end{equation}
The angular centroid on the other hand is the corresponding centroid of the transverse Fourier transform of the field profile
\begin{equation}
    \mathcal{E}(\phi_x,\phi_y,z) = \int E(x,y,z)e^{-ik_r(\phi_xx+\phi_yy)}dxdy.
\end{equation}
It is readily found that the centroid of $|\mathcal{E}(\phi_x,\phi_y,z)|^2$ is
\begin{equation}
    \phi_{x,cen} = -\frac{\left|Q_x\right|^2\Im[x_0]}{k_r\Im[Q_x]}.
\end{equation}
It is also worth noting that we can invert these centroids to write the original complex centroid parameter in terms of the physical and angular centroids
\begin{equation}
    x_0 = x_{cen}+\frac{k_r\phi_{x,cen}}{Q_x}.
\end{equation}
Similarly, the $Q$ parameters can be written in terms of the rms size of the intensity if it comes to a waist, $\sigma_w$, as well as the distance from the current position to the waist,  $z_w$. The expression is 
\begin{equation}
    Q = -\frac{ik_r}{2k_r\sigma_w^2+iz_w}.
\end{equation}
Finally, the time-averaged radiation power is the usual integral over the transverse intensity profile
\begin{eqnarray}
    P(z) =&& \frac{\epsilon_0c}{2}\int \left|E(x,y,z)\right|^2dxdy\\
    =&&\sigma_x\sigma_y\pi\epsilon_0c\left|f(z)\right|^2e^{-\frac{|Q_x|^2\Im[x_0]}{\Im[Q_x]}-\frac{|Q_y|^2\Im[y_0]}{\Im[Q_y]}}.
\end{eqnarray}
We note that although we call this power, it should more accurately be understood as spectral power since we consider a single frequency slice of the radiation field.

\section{\label{app:index}Analytic expression for the FEL refractive index with a gaussian mode}

In this section we will give the analytic form of the FEL refractive index for a gaussian mode and arbitrary electron trajectory. Since the impact of electron trajectory jitter is in some sense an additional impact on top of x-ray trajectory offsets, in particular in the case of the XRAFEL, we will explicitly separate out their impacts. The end result is of the form
\begin{widetext}
\begin{equation}
    n^2(x,y,z,\zeta) = 1+\frac{4\pi k_\beta^2\sigma_x^2}{k_r}\int_0^zd\zeta\frac{f(\zeta)}{f(z)}K_{10}(z,\zeta)\sqrt{\frac{1}{g_x(z,\zeta)g_y(z,\zeta)}}e^{h_{0}(x,z,\zeta)+h_{ce}(x,z,\zeta)+h_{0}(y,z,\zeta)+h_{ce}(y,z,\zeta)},
\end{equation}
\end{widetext}
where $g_x(z,\zeta)=i+k_rk_\beta^2\sigma_x^2(z-\zeta)-\sigma_x^2Q_x(\zeta)\sin(k_\beta(z-\zeta))^2$ with a similar expression for y. The arguments of the exponential take the form
\begin{widetext}
\begin{eqnarray}
    h_{0}(x,z,\zeta) =&& \frac{i}{2}\left[ Q_x(z)(x-x_0(z))^2 + \frac{A(z,\zeta)\left[x^2A(z,\zeta)-\sigma_x^2Q_x(\zeta)(x^2-2xx_0(\zeta)\cos(k_\beta(z-\zeta))+x_0(\zeta)^2)\right]}{\sigma_x^2\left(A(z,\zeta)-\sigma_x^2Q_x(\zeta)\sin(k_\beta(z-\zeta))^2\right)} \right],\\
    \nonumber h_{ce}(x,z,\zeta) =&& \frac{p_{x,ce}(z)\left[ip_{x,ce}(z)+2k_\beta\sigma_x^2Q_x(\zeta)\sin(k_\beta(z-\zeta))\left(x_0(\zeta)-x\cos(k_\beta(z-\zeta))\right)\right]}{2k_\beta^2\sigma_x^2\left(A(z,\zeta)-\sigma_x^2Q_x(\zeta)\sin(k_\beta(z-\zeta))^2\right)}\\
    &&-\frac{p_{x,ce}(z)^2}{2k_\beta^2\sigma_x^2}-\frac{x_{ce}(z)^2-2xx_{ce}(z)}{2\sigma_x^2} ,
\end{eqnarray}
\end{widetext}
where in each of these $A(z,\zeta)=i+k_rk_\beta^2\sigma_x^2(z-\zeta)$. In these forms, $h_{ce}$ vanishes in the absence of a non-zero electron beam trajectory. When $x_0(z)$ and $y_0(z)$ also vanish, the ideal FEL scenario without x-ray offset is retrieved. 

\bibliography{apssamp}

\end{document}